\documentclass[lettersize,journal]{IEEEtran}

\usepackage{amsmath,amsfonts}
\usepackage{algorithmic}
\usepackage{algorithm}
\usepackage{array}
\usepackage[caption=false,font=normalsize,labelfont=sf,textfont=sf]{subfig}
\usepackage{textcomp}
\usepackage{stfloats}
\usepackage{url}
\usepackage{verbatim}
\usepackage{graphicx}
\usepackage{cite}
\usepackage[inline]{enumitem}
\usepackage{cleveref}
\usepackage[separate-uncertainty = true,multi-part-units=single]{siunitx}
\usepackage{booktabs}
\usepackage{subcaption}
\usepackage{xcolor}
\usepackage{pifont}
\newcommand{\cmark}{\ding{51}}%
\newcommand{\xmark}{\ding{55}}%
\usepackage{tikz}
\usepackage{eso-pic}

\begin{document}

\title{A Non-Invasive Path to Animal Welfare: \\ Contactless Vital Signs and Activity Monitoring of In-Vivo Rodents Using a mm‐Wave FMCW Radar} % temporary title 

\author{Tommaso~Polonelli,~\IEEEmembership{Senior Member,~IEEE}, Manuel Glahn, Stefano Kron, Stefan Selbert, Marco Garzola, and Michele~Magno,~\IEEEmembership{Senior Member,~IEEE}
        % <-this % stops a space
\thanks{T. Polonelli, M. Glahn, S. Kron and M. Magno are with the Center for Project-Based Learning of ETH Z\"urich, ETZ, Z\"urich, Switzerland (e-mail: tommaso.polonelli@pbl.ee.ethz.ch). S. Selbert is with the Phenomics Center (EPIC) ETH Z\"urich, Z\"urich, Switzerland. M. Garzola is with TECNIPLAST S.p.A., Italy (e-mail: marco.garzola@tecniplast.it).}%
\thanks{This work is supported by TECNIPLAST S.p.A., Italy, which provided the GM500 home cage and technical support. We also thank Johannes Promny  Torsten for his valuable support.}% <-this % stops a space
}

\AddToShipoutPictureFG*{%
  \AtPageUpperLeft{%
    \begin{tikzpicture}[remember picture,overlay]
      \node[anchor=north] at ([yshift=-2mm]current page.north){%
        \parbox{\paperwidth}{\centering \color{gray}\bfseries
          This work has been submitted to IEEE for possible publication. \\ Copyright may be transferred without notice, \\ after which this version may no longer be accessible.%
        }%
      };
    \end{tikzpicture}
  }
}

%\markboth{Journal of \LaTeX\ Class Files,~Vol.~14, No.~8, August~2021}{Shell \MakeLowercase{\textit{et al.}}: A Sample Article Using IEEEtran.cls for IEEE Journals}


\maketitle

%---------------------------------------------------------------------------------------------------
%TODO! find a nice catchy name! like: AIglass Capacitive eye tracking  Wearable Non-contact, ceog, etc.
% let's also try to find a catchier first sentence
\begin{abstract}
Monitoring physiological and behavioral parameters of laboratory rodents is fundamental for biomedical research, yet conventional techniques often rely on invasive sensors or frequent handling that can induce stress and compromise data fidelity. To address these limitations, this paper presents a  contactless and non-invasive in-vivo monitoring system based on a low-power 60\,GHz frequency-modulated continuous wave (FMCW) radar. The proposed system enables simultaneous detection of rodent activity and vital signs directly within home-cage environments, eliminating the need for implants, electrodes, or human intervention. 
The hardware platform leverages a compact Infineon BGT60 series radar sensor, optimized for low power consumption and continuous operation. We investigate sensor placement strategies and design a complete signal processing pipeline, including range bin selection, phase extraction, and frequency-domain estimation tailored to rodent vital signs. The system achieves 3~cm and 0.1\,$\mathbf{ms^{-1}}$ sensitivity for motion and activity detection, while allowing discrimination of micro-movements associated with cardiopulmonary activity with a 2\,$\mathbf{\mu m}$ distance resolution. Experimental validation with two rodents in realistic in-vivo cages demonstrates that the radar can track animal position and extract respiration rates with $\mathbf{\pm}$2\,bpm accuracy. By minimizing stress and disturbance, this work improves both animal welfare and the reliability of physiological measurements, offering a refined alternative to traditional monitoring methods. This work represents the first demonstration of continuous radar-based vital sign monitoring in freely moving rodents within group-housed cages. The proposed approach lays the foundation for scalable, automated, and ethical monitoring solutions in preclinical and translational research. 
\end{abstract}


%---------------------------------------------------------------------------------------------------
\section{Introduction}
\label{sec:intro}
Laboratory rodents play a critical role in biomedical research, but continuously monitoring their physiological signals in-vivo presents significant challenges \cite{bourquin_vivo_2022, fusco_deep_2025}. Traditional methods for measuring vital signs in small animals often rely on implantable telemetry or contact sensors (e.g., ECG electrodes or PPG probes) that require restraining or surgically instrumenting the animal~\cite{sun_contactless_2025}. Such invasive approaches not only pose ethical concerns but can themselves induce stress and alter the very parameters being measured~\cite{sun_contactless_2025,zhang_contactless_2024}. Even handling or brief restraint (for example, simply opening the cage to attach sensors) can cause distress \cite{hurst_taming_2010, balcombe_laboratory_2004}, potentially impacting experimental results \cite{panzacchi_carcinogenic_2025}. This has created a demand for non-invasive, stress-free monitoring techniques, whereby animal welfare is improved without compromising scientific data quality \cite{balcombe_laboratory_2004,polonelli_multi-protocol_2019, hu_novel_2024}. Indeed, recent studies highlight the need for simple, contactless methods to collect heart rate (HR) \cite{hu_novel_2024} and respiration rate (RR) \cite{fusco_multirange_2025} data from animals without training or intrusive equipment~\cite{zhang_contactless_2024}. 

Concurrently, a research trend has emerged toward contactless vital sign monitoring across diverse fields ranging from human healthcare to veterinary science~\cite{singh_multi-resident_2021}. Various remote sensing technologies – including camera-based systems \cite{hu_novel_2024}, thermal imaging, digital ventilated cage (DVC) \cite{fuochi_phenotyping_2021, iannello_non-intrusive_2019}, and Doppler radars \cite{xiong_human_2025} – have been explored to detect physiological signals without direct contact \cite{rong_noncontact_2021}. Optical methods (e.g., RGB cameras) can track breathing via body motions or subtle color changes, but they are sensitive to lighting, temperature, and an animal’s fur or coloration~\cite{singh_multi-resident_2021}. In contrast, millimeter-wave (mmWave) radars \cite{banerjee_efficient_2025,will_human_2019} have gained attention as a robust solution that can penetrate materials like clothing or bedding with minimal attenuation~\cite{marty_frequency_2024}. Radar systems measure micrometric movements of the body surface caused by breathing and cardiac activity, offering sub-millimeter precision in displacement detection~\cite{zhang_photonic_2023}. Recent advances in low-power radar technology have enabled compact, battery-operable devices suitable for continuous monitoring applications~\cite{marty_investigation_2023, fusco_deep_2025}. Studies on humans have already demonstrated that mmWave frequency-modulated continuous wave (FMCW) radars can accurately capture heart and respiration signals without contact, even through intervening materials, and from a distance~\cite{marty_frequency_2024,marty_investigation_2023}. In particular, comparative evaluations of \qty{24}{\giga\hertz},  \qty{60}{\giga\hertz}, and  \qty{120}{\giga\hertz} FMCW radars showed that all three could detect human cardiac and respiratory activity (with higher frequencies like \qty{120}{\giga\hertz} providing the strongest performance)~\cite{marty_frequency_2024}. These advancements underscore the potential of radar-based approaches for unobtrusive vital sign monitoring in new domains, including animal research~\cite{zhang_photonic_2023,sun_contactless_2025}. 

Translating radar monitoring to small animal subjects introduces additional considerations and challenges. Prior works have begun to investigate radar for animal physiology, but mostly in controlled settings \cite{zhang_photonic_2023,sun_contactless_2025,singh_multi-resident_2021}. For example, an mmWave FMCW radar placed against the enclosure was used to monitor heart and breathing rates in anesthetized macaque monkeys, achieving near real-time estimation of HR and RR with errors under \qty{1}{bpm} and \qty{1.3}{rpm}, respectively~\cite{zhang_photonic_2023}. Likewise, a compact \qty{24}{\giga\hertz} radar has been applied to track respiratory rate variability in anesthetized rats, demonstrating high accuracy while minimizing animal handling~\cite{sun_contactless_2025}. These preliminary studies validate the feasibility of contactless vital sign detection in animals and emphasize its benefits for animal welfare (e.g., avoiding implants or restraint)~\cite{sun_contactless_2025}. On the other hand, this work did overcome the challenge to extend such techniques to awake, freely moving rodents in their normal housing environment. In a free-behavior scenario with group-housed mice, the radar copes with motion, multiple targets, and smaller signal amplitudes due to the animals’ size and rapid physiology. To date, few established solutions exist for non-invasive monitoring of unrestrained rodents \cite{sturman_grimace_2025}; conventional approaches either require isolating the subject or forgoing vital sign measurement in favor of simpler activity monitoring~\cite{singh_multi-resident_2021}. This gap motivates our work to pioneering the development of a radar-based system specifically tailored to monitor vital signs in lively, group-housed rodent environments.

This paper presents an accurate contactless vital-sign monitoring system for rodents, enabling continuous in-cage tracking of activity and respiration without the need for implanted or wearable sensors. The proposed approach employs a low-power \qty{60}{\giga\hertz} FMCW radar sensor (Infineon BGT60 series) positioned optimally with respect to the rodent’s home cage to capture subtle chest movements associated with breathing. The system detects the presence and location of a target mouse and measure the breathing rate in real-time. We also investigate the potential to extract heart rate from the radar micro-movements, leveraging the \qty{60}{\giga\hertz} carrier frequency and \qty{5}{\giga\hertz} bandwidth to maximize sensitivity to the  $\sim$\qty{4}{\micro\meter} cardiac activity. By covering the full implementation from the sensor hardware up to the data processing algorithms, our work addresses practical considerations such as optimal sensor placement (e.g., below the bedding or mounted under the cage for best line-of-sight to the animal) and signal processing techniques tuned to rodent vital signatures. The resulting system operates entirely non-invasively – no electrodes, implants, or even cage openings are required – thereby significantly reducing animal stress and improving the authenticity of the collected physiological data. This refinement in experimental technique not only aligns with ethical standards (3R\footnote{The \textit{3Rs} principles are Replace, Reduce, and Refine, a set of ethical guidelines for animal research that aims to minimize animal use and suffering.} principles in in-vivo trials \cite{mosch_towards_2023}, reducing human intervention to a minimum) but also enhances scientific outcomes by allowing rodents to behave naturally during monitoring thereby improving experimental reproducibility (4R \cite{mosch_towards_2023}). Indeed, findings from $80+$ studies \cite{balcombe_laboratory_2004} and from Giovannini \textit{et al.} \cite{panzacchi_carcinogenic_2025} show that animals exhibit rapid, pronounced, and statistically significant increases in stress-related responses during all in-vivo tested procedures. Deviations from baseline or control values generally ranged between 20\% and over 100\% \cite{balcombe_laboratory_2004}, persisting for at least 30 minutes or more. These results suggest that standard laboratory practices induce stress in animals and that they do not easily habituate to such routines \cite{balcombe_laboratory_2004}. Overall, the evidence indicates that considerable fear, stress, and potentially distress are predictable outcomes of routine laboratory procedures, carrying important scientific and ethical implications for animal research \cite{balcombe_laboratory_2004}.

The contributions of this work can be summarized as follows:
\begin{enumerate*}[label=(\roman*),font=\itshape]
    \item Contactless in-vivo rodent monitoring system: we design and implement a radar-based monitoring system that track rodent activities with \qty{3}{\cm} resolution and measure respiration in real time without any contact reaching a 99\% accuracy compared to a image-based ground-truth. The system, depicted in \Cref{fig:overview}, enables non-invasive breathing measurement in freely moving mice. This is a pioneering demonstration for vital sign monitoring in group-housed rodents within their home cage environment.
    \item Optimized sensing and signal processing pipeline: the paper illustrates a comprehensive pipeline from hardware to data analysis, including optimal sensor placement and orientation for in-cage monitoring, and custom signal processing algorithms for vital sign extraction. This pipeline addresses the unique challenges of rodent monitoring – such as high respiratory rates \qtyrange{150}{300}{bpm}, small cardiac motion amplitude \qtyrange{4}{32.4}{\micro\meter}, and interference from cage mates or environment – by combining deterministic filtering and spectral techniques to isolate respiration-induced movements. 
    \item Experimental validation and welfare impact: we validate the proposed system in realistic conditions with  two female Swiss Webster mice, demonstrating successful detection of a target animal’s breathing rate and activity levels without any human handling. The results show that our contactless method can reliably (99\% accuracy) monitor respiratory trends and detect movements of rodents even during normal cage activities.
\end{enumerate*}

\section{Related Works}
\label{sec:rel-works}
The push for measuring vital signs of living subjects and detect their position with a contactless and non-invasive solution is a megatrend in research \cite{hui_no-touch_2019, li_metaphys_2023}, including bio-engineer \cite{marty_frequency_2024}, medicine \cite{ruiz-zafra_neocam_2023}, veterinary (enforced by country-specific laws) \cite{kruljac_challenges_2024}, IoT smart-buildings \cite{rzucidlo_non-invasive_2023} and recent vehicles \cite{blangiardi_exploring_2024}. Recently, with the advent of novel sensing techniques and signal processing, non-invasive and contactless vital sign and localization monitoring systems are appearing in the research and industrial field \cite{luo_applications_2024}. With human subjects, solutions are mainly related with real-time monitoring during hospitalization or medical cares, mainly based on cameras \cite{ruiz-zafra_neocam_2023}, while for veterinary and animals the application scenarios are wider \cite{darlis_autonomous_2023}. For example, there is a push for the 3R principles in in-vivo trials, or for animal tracking in zoos, or to decrease the animal stress during veterinary visits \cite{fuochi_phenotyping_2021}. Recent works are proposing interesting solutions for non-invasive tracking, but are mainly limited to only tracking activity/behavior \cite{pitafi_cagedot_2024, rodina_automated_2022} or estimating vital signs \cite{zhang_contactless_2024}. Among different sensing techniques, the most common are summarized in \Cref{tab:soa-comparison}, including vision-based systems, radars, vibration, microphones and thermal sensing \cite{kwon_non-contact_2021}. As shown in \Cref{tab:soa-comparison}, no previous studies have explored the use of FMCW radars with freely moving rodents in uncontrolled in-vivo environments — the focus of the present work.
\begin{table*}[t]
    \caption{Comparison with state-of-the-art (SoA) of non-invasive contactless vital sign measurement and activity-movement classification. The table summarizes the main supported features of each reference work with a qualitative metric.}
    \label{tab:soa-comparison}
    \centering
    \footnotesize
    \renewcommand{\arraystretch}{1.2}
    \begin{tabular}{c c c c c c c c c c}
        \hline
        \hline
        %\rowcolor[HTML]{E0E0E0} 
        \begin{tabular}[c]{@{}c@{}} 
            Reference \\ work \end{tabular} & \begin{tabular}[c]{@{}c@{}}Sensing \\
            method\end{tabular} &  \begin{tabular}[c]{@{}c@{}} 
            Non \\ invasive \end{tabular}  & \begin{tabular}[c]{@{}c@{}}Movement\\
            detection$^{\star\star}$\end{tabular} & Localization & RR & HR & \begin{tabular}[c]{@{}c@{}}Multiple \\
             organisms \end{tabular} & \begin{tabular}[c]{@{}c@{}} Static \\
             Dynamic$^\star$ \end{tabular} & \begin{tabular}[c]{@{}c@{}}Organism  
        \end{tabular}  \\
        \hline  
        \cite{li_contactless_2024} & FMCW \qty{77}{\giga\hertz} radar  & \cmark & \cmark & \cmark & \cmark & \cmark & \xmark & D & Human \\ 
        \cite{yao_noncontact_2024} & FMCW \qty{77}{\giga\hertz} radar  & \cmark & \xmark & \xmark & \cmark & \cmark & \xmark & S & Human \\ 
        \cite{ruiz-zafra_neocam_2023} & RGB Camera  & \cmark & \cmark & \xmark & \cmark & \xmark & \xmark & S & Human \\ 
        \cite{marty_investigation_2023} & FMCW \qty{24}{\giga\hertz} radar  & \cmark & \xmark & \xmark & \cmark & \cmark & \xmark & S & Phantom \\ 
        \cite{blangiardi_exploring_2024} & IR-UWB  & \cmark & \cmark$^\diamond$ & \xmark & \xmark & \xmark & \cmark & S & Human \\ 
        \cite{pitafi_cagedot_2024} & Geophone sensors  & \cmark & \cmark & \xmark & \xmark & \xmark & \cmark & D & Rodent \\ 
        \cite{rodina_automated_2022} & \qty{24}{\giga\hertz} radar  & \cmark & \cmark & \xmark & \xmark & \xmark & \xmark & D & Rodent \\ 
        \cite{kwon_non-contact_2021} & Thermography camera  & \cmark & \xmark & \xmark & \xmark & \cmark & \xmark & S & Human \\ 
        \cite{zhang_contactless_2024} & FMCW \qty{64}{\giga\hertz} radar  & \xmark & \xmark & \xmark & \cmark & \cmark & \xmark & S & Macaque \\ 
        \cite{iannello_non-intrusive_2019} & Capacitive  & \cmark & \cmark & \cmark & \xmark & \xmark & \xmark & D & Rodent \\ 
        Our work & FMCW \qty{60}{\giga\hertz} radar  & \cmark & \cmark & \cmark & \cmark & \xmark & \cmark & D & Rodent \\ 
        \hline
        \hline
        \multicolumn{10}{l}{$^\star$Static includes all situations in which the subject need to be static or in a specific position, Dynamic includes free and uncontrolled movements,} \\
        \multicolumn{10}{l}{$^{\star\star}$Includes also activity classification, $^\diamond$People counting.} 
    \end{tabular}
\end{table*}
The study in \cite{zhang_contactless_2024} introduces an innovative approach to non-invasive monitoring of HR and RR in non-human primates (NHPs), specifically macaques. Utilizing a FMCW radar system, the researchers positioned the device to face the chest of both awake and anesthetized macaques within their enclosures. This configuration enabled real-time estimation of HR and RR without direct contact or the need for restraint, thereby minimizing stress-induced alterations in physiological parameters. The study demonstrates the feasibility of using mm-wave FMCW radar technology for contactless monitoring of vital signs in NHPs, a method previously underexplored in this context. However, the study's findings are based on a specific group of macaques, and a related with controlled laboratory conditions, which may limit the result generalization across different NHP species or individuals with varying health conditions. Factors such as enclosure materials, positioning of the radar, and movement of the animals could affect the accuracy and reliability of the measurements, necessitating further research to optimize the system for diverse settings. In a similar direction, authors in \cite{marty_investigation_2023} examines the use of mmWave radars for contactless monitoring of vital signs. The study focuses on three FMCW low-power radars operating at carrier frequencies of \qtyrange{24}{120}{\giga\hertz}. Data were collected from human subjects to determine the feasibility of detecting heartbeats and breathing patterns using simple algorithms suitable for low-power embedded processors. All three radars successfully identified cardiac and respiratory activities, with the \qty{120}{\giga\hertz} system demonstrating superior performance. However, the study acknowledges limitations, including the controlled environment of the experiments, which may not fully represent real-world conditions where factors such as movement and varying distances could affect performance. The paper \cite{yao_noncontact_2024} optimizes the approach investigated in \cite{marty_investigation_2023, zhang_contactless_2024, li_contactless_2024} with an efficient and robust algorithm for estimating vital signs using a FMCW radar. The authors derive a maximum likelihood estimator based on Newton's method to accurately assess breathing and heart rates. However, the study acknowledges certain limitations. The performance of the proposed algorithm may be affected by factors such as subject movement and varying environmental conditions, which can introduce noise and artifacts into the radar signals. Therefore, although explored in the literature, FMCW radar are so far tested mainly in controlled and static conditions, and never fully evaluated with free animals during normal daily activities.

As evident from \Cref{tab:soa-comparison}, our work is the only one addressing localization and vita-sign monitoring in a dynamic and uncontrolled scenario with multiple animals. Our results, conducted in-vivo with multiple rodents, confirm the possibility to detect the animal activity, its position in the cage, while detecting in real-time the respiration rate. %Our system relies on a single low-power FMCW \qty{60}{\giga\hertz} radar combined with lightweight data processing. 

\section{Methodology}

\subsection{Radar Background}
\label{sec:radarback}
Radars operate based on the illuminate-echo principle, in which the system first transmits an electromagnetic wave that propagates through space and is subsequently reflected by an object \cite{zhang_contactless_2024}. The reflected signal returns to the radar after a propagation delay, providing information about the target. The properties of the transmitted signal determine the type of radar and the characteristics it can measure. FMCW radars employ a frequency-modulated signal, typically centered around a carrier frequency \( f_c \). Various modulation schemes exist, such as triangular and sinusoidal, but this work primarily focuses on saw-tooth modulation, which linearly sweeps the frequency from \( f_{\text{min}} \) to \( f_{\text{max}} \)\footnote{A complete explanation of FMCW radars falls outside the scope of this paper; therefore, we point the reader to \cite{marty_frequency_2024} for a deeper explanation.}. The bandwidth \( B \) is defined as the difference between these frequencies, while the modulation slope \( S \) characterizes the rate of frequency change. In the time domain, the transmitted signal, known as a chirp, is expressed in \Cref{eq:timedomain}, where \( A_{TX} \) is the amplitude of the transmitted signal, $t$ the time, and \( T_c \) represents the chirp duration. When an object is located at a distance \( d \), the reflected signal experiences a round-trip delay \( \tau = \frac{2d}{c} \), where \( c \) denotes the speed of light. This delay is represented in \Cref{eq:timedelay}, where \( A_{\text{IF}} \) is the amplitude of the received signal, $\phi$ the phase, and \( \lambda \) is the wavelength of the transmitted signal. This expression can be rewritten in terms of the intermediate frequency (IF) signal, as in \Cref{eq:intermediate}.
\begin{equation}
x_T(t) = A_{TX} \cos \left( 2\pi f_{\text{min}} t + \pi \frac{B}{T_c} t^2 \right)~,
\label{eq:timedomain}
\end{equation}
\begin{equation}
S_{\text{IF}}(t) \approx A_{\text{IF}} \cos \left( 2\pi \frac{B}{T_c} \tau t + 2\pi \frac{2d}{\lambda} \right)~,
\label{eq:timedelay}
\end{equation}
\begin{equation}
S_{\text{IF}} = A_{\text{IF}} \cos (2\pi f_{\text{IF}} t + \phi), \quad f_{\text{IF}} = \frac{B}{T_c} \frac{2d}{c}~.
\label{eq:intermediate}
\end{equation}
From the above equation, it is evident that the IF frequency \( f_{\text{IF}} \) is directly proportional to the target distance \( d \), depicted in \Cref{fig:overview}, and the bandwidth $B$. By performing a frequency analysis of the IF signal using the Fast Fourier Transform (FFT), the target distance can be accurately determined. In real-world scenarios, reflections from surrounding objects and noise add additional complexity to the received signal, which must be accounted for processing. For commercial mmWave radars, the range resolution ($d_{\text{res}}$ in \Cref{eq:submillimeterdisplacements}) is typically in the order of centimeters, which is insufficient for precise displacement measurements required for vital sign monitoring. However, by analyzing the phase shift between consecutive chirps, sub-millimeter displacements can be estimated using the relation described in \Cref{eq:submillimeterdisplacements}, where \( \Delta \phi \) denotes the phase difference between successive chirps. This phase-based approach allows FMCW radars to achieve high sensitivity in tracking minute displacements, making them suitable for applications such as contactless vital sign monitoring.
\begin{equation}
d_{\text{res}} = \frac{c}{2B},~\Delta d = \frac{\lambda \Delta \phi}{4\pi}
\label{eq:submillimeterdisplacements}
\end{equation}

Theoretically, estimating the distance to an object requires transmitting a single chirp. However, velocity estimation necessitates the transmission of at least two chirps. When the radar signal reflects off a moving object, the received chirps exhibit similar amplitudes but distinct phase differences. This phase shift is crucial for determining the object's velocity. Given \( N_c \) chirps in a frame, applying a FFT to each chirp produces a range-FFT, which identifies peaks corresponding to the object's position. These peaks appear at the same range location but with varying phases.

The phase difference \( \omega \) in \Cref{eq:omega} observed between consecutive chirps can be used to estimate velocity, where \( v \) represents the object's velocity, \( \lambda \) denotes the signal wavelength, and \( T_{c,r} = T_{c} + T_{c,i}\) is the chirp repetition time, which consists of \(T_{c}\) and the chirp idle time \(T_{c,i}\). The method is effective as long as the phase shift remains within \( |\omega| < \pi \) (i.e., \ang{180}). If the phase shift exceeds this limit, ambiguity arises, making it challenging to differentiate between positive and negative phase shifts, which introduces an upper bound on the measurable velocity in \Cref{eq:omega}, where \( v \) represents the object's velocity, \( \lambda \) denotes the signal wavelength, and \( T_c \) is the chirp interval.
\begin{equation}
\omega = \frac{4\pi v T_{c,r}}{\lambda}  \Rightarrow  v = \frac{\lambda \omega}{4\pi T_{c,r}} \Rightarrow v_{\text{max}} < \frac{\lambda}{4T_{c,r}}
\label{eq:omega}
\end{equation}
When monitoring multiple objects at the same range but with different velocities, the use of more than two chirps is necessary. A sequence of \( N_c \) equally spaced chirps constitutes a frame. After performing the range-FFT, a second FFT is applied across the chirps, resulting in the so-called Doppler FFT. The Doppler FFT output is a two-dimensional matrix, also called Range-Doppler map, containing both range and velocity information, enabling the radar to distinguish multiple objects with varying velocities at different ranges. Considering two different phase shifts between consecutive chirps, \( v_1 = \frac{\lambda \omega_1}{4\pi T_{c,r}} \) and \( v_2 = \frac{\lambda \omega_2}{4\pi T_{c,r}} \), and using the condition that two frequencies \( \omega_1 \) and \( \omega_2 \) can be resolved in an FFT of length \( N_c \) if \( |\omega_1 - \omega_2| > \frac{2\pi}{N_c} \), the velocity resolution \( \Delta v \) can be expressed in \Cref{eq:deltaomega}.
\begin{equation}
\Delta \omega = \frac{4\pi \Delta v T_{c,r}}{\lambda}, \quad \Delta \omega > \frac{2\pi}{N_c} \quad \Rightarrow \quad \Delta v > \frac{\lambda}{2N_c T_{c,r}}
\label{eq:deltaomega}
\end{equation}
Thus, improving velocity resolution requires increasing the chirp repetition time or the number of chirps per frame. The total frame time \( T_f = N_{c} * T_{c,r} + T_{f,i}\) consists additionally of the frame idle time \(T_{f,i}\). Most of the time, the idle frame is much smaller than the active frame time, in this case the velocity resolution, \( v_{\text{res}} \), is inversely proportional to the total active time, as expressed in \Cref{eq:vres}.
\begin{equation}
v_{\text{res}} \cong \frac{\lambda}{2 T_f}, \text{  if  } f_{i} << N_{c} * T_{c,r}
\label{eq:vres}
\end{equation}
\subsection{Heavy Clutter Environments}\label{sec:cage_env}
A major challenge when working with FMCW radars is dealing with multipath reflections and occlusion. A rodent cage is considered a heavy-clutter environment, see \Cref{fig:Höngg_experiment_setup}, as it contains various reflective components both inside and around the cage, such as plastic cage enrichment, plastic walls, and metallic food or water containers. These static components interact with the moving animals, creating temporary multipath and shadow ghosts with comparable signal-to-noise ratios masking the target signal \cite{chen_environment-aware_2023}.
There are concepts to compensate for multipath and shadowing ghost effects, such as reflection maps. For simplicity, the authors of this paper employed a running average filter (RAF) to eliminate stationary clutter sources, as suggested by \cite{kim_robust_2020}.
The RAF enables dynamic filtering of artifacts introduced by static reflections or shadowing effects. The parameter $\alpha$ of the RAF can be adjusted to optimally filter out the artifacts while keeping the movement information of the mouse.

\subsection{Algorithms and Digital Signal Processing}
\label{sec:dsp}
Extracting vital signs, estimating the level of movement of mice, and determining distance information using radar require different signal processing approaches.

For vital sign processing, the tiny oscillations corresponding to respiration and heart rate can be detected by extracting the phase. First, to accurately assess target displacement at the correct distance, the appropriate range bin must be identified. The range bins have a resolution of $d_{res}$, expressed in \Cref{eq:submillimeterdisplacements}. Subsequently, HR and RR are derived from the phase or displacement signal. This phase approach has a significant advantage, it allows for the derivation of individual breathing patterns, whether for a single mouse or multiple mice, under the assumption that the mice are distributed across different range bins. Although several algorithmic methods, this study employs a deterministic algorithm chain to enhance deployability in different application scenarios, a common limitation of fully data-driven learning-based approaches. Each stage of the algorithm is tailored to meet the strict resource constraints of embedded processors with limited memory and computational capacity, ensuring suitability for deployment on battery-operated mobile devices or systems where processing must occur in close proximity to the sensors. The processing pipeline for vital sign monitoring subsequently performs chirp accumulation, range bin selection, phase extraction, and ultimately, vital sign estimation.

To enhance the signal-to-noise ratio (SNR), 16 chirps (from \Cref{sec:radarback}) are transmitted sequentially in a brief burst lasting \qty{1.2}{\milli\second}. Assuming the target remains stationary during this interval, the effective SNR of the phase can be improved by increasing the signal’s integration time via averaging. Although this method results in higher data throughput and increased power consumption, it does not compromise the subsequent memory and computational efficiency since the averaging is performed immediately after data acquisition. After performing the range FFT on the averaged chirps, the next step is to identify the subject by selecting the appropriate range bin, known as the target bin. To avoid stationary objects (e.g. metallic food container) from being selected as target bin, the target bin is selected as the one exhibiting the highest magnitude variance. Additionally, the range bins corresponding to the first \qty{5}{\cm} are not taken into account since Tx-Rx leakage is present in these bins. In a next step, the respiration signal is derived by extracting the phase of the target bin over the slow-time. 

Since respiration can vary over time and is not necessarily strictly periodic, we adopt a time-domain approach for its estimation. This method is feasible because the displacement due to breathing is typically one to two orders of magnitude larger than that caused by cardiac activity \cite{marty_frequency_2024}. Initially, the signal is filtered within the range corresponding to typical breathing rates (detailed in \Cref{sec:mousephysiology}), yielding a smooth waveform on which local maxima can be reliably identified. The median interval between these peaks is then computed to estimate the breathing frequency. This methodology was consistently applied across all three radar antennas to generate robust RR estimates. 

Time-domain methods, such as the one employed for RR, proved to be less reliable for HR estimation due to the low SNR associated with cardiac-induced displacements. Consequently, frequency-domain techniques are favored for HR estimation. Therefore, our primary observation is that, while breathing patterns exhibit larger amplitude displacements, the cardiac component is characterized by a higher acceleration due to impulsive muscular contractions, as also observed in \cite{chen_contactless_2024}. Accordingly, we derive the chest acceleration as an intermediate signal from the displacement data. This acceleration is computed using the second-order derivative coefficients of a Savitzky–Golay filter with a seven-sample window (\qty{70}{\milli\second}) and a third-order polynomial. Since motion direction is irrelevant for HR estimation, the absolute envelope of the acceleration is obtained by summing the positive and negative components, which effectively suppresses breathing harmonics. The HR is then estimated in the frequency domain, providing additional robustness against false or missing heartbeats resulting from the low SNR of cardiac activity. Zero padding in the FFT facilitates fine-grained peak localization in the frequency spectrum.  

The processing pipelines for position estimation and activity classification share a similar structure, both relying on the range-Doppler map, which provides range and velocity information. Before deriving the range-Doppler map, the radar data undergoes preprocessing. First, mean removal is applied to emphasize relative changes rather than absolute radar signal values. Next, the RAF is implemented to mitigate static reflections and shadowing effects caused by the cage walls and objects inside the cage. The range-Doppler map is then computed using zero-padding for both the range FFT (number of samples) and the Doppler FFT (number of chirps), with a Hanning window applied in both cases. Subsequently, for distance estimation, the L2 norm of all velocity entries per range bin is calculated, reducing the two-dimensional range-Doppler map to a one-dimensional range-power array. Each range bin now represents the power of movement for the corresponding distance. The range bin with the highest movement power values is identified as the location of the mouse. To estimate the overall movement level of the mouse, the total movement within the cage is determined by flattening the range-Doppler map and applying the L2 norm on the resulting array. The bins in the range-Doppler map corresponding to the first \qty{5}{\cm} are clipped away since Tx-Rx leakage is present in these bins.
For distance estimation, we derive the power of all velocity bins in the range-Doppler map, determining the movement level at each range. Identifying the range bin with the highest movement level provides the distance to the mouse inside the cage. In contrast, for overall movement level estimation within the cage, the power is not only derived from the velocity bins but from the entire range-Doppler map. This approach quantifies the total movement power inside the cage.

%Since this paper serves as a pioneering study on the feasibility of rodent monitoring in a dynamic and uncontrolled scenario using FMCW radar, the signal processing approaches in this chapter were primarily designed for single-mouse monitoring. This allows us to initially approach the problem in the simplest way before gradually increasing complexity by incorporating multiple mice. The goal is to first assess feasibility and evaluate radar performance in single-mouse experiments, followed by an investigation into the feasibility of multi-mouse experiments.
%It must be considered that even with just two mice, the problem complexity increases drastically compared to the single-mouse case. For example, the movement level approach measures the total movement occurring within the cage. With a single mouse, it is clear that all detected movement originates from that one mouse. However, with two mice, it becomes uncertain whether the implemented movement level approach can distinguish between two mice moving simultaneously and one mouse moving while the other remains stationary (as would be the case with a dead mouse). 

\section{System Configuration and Validation}

\begin{figure}[t]
    \centering
    \includegraphics[width=0.99\columnwidth]{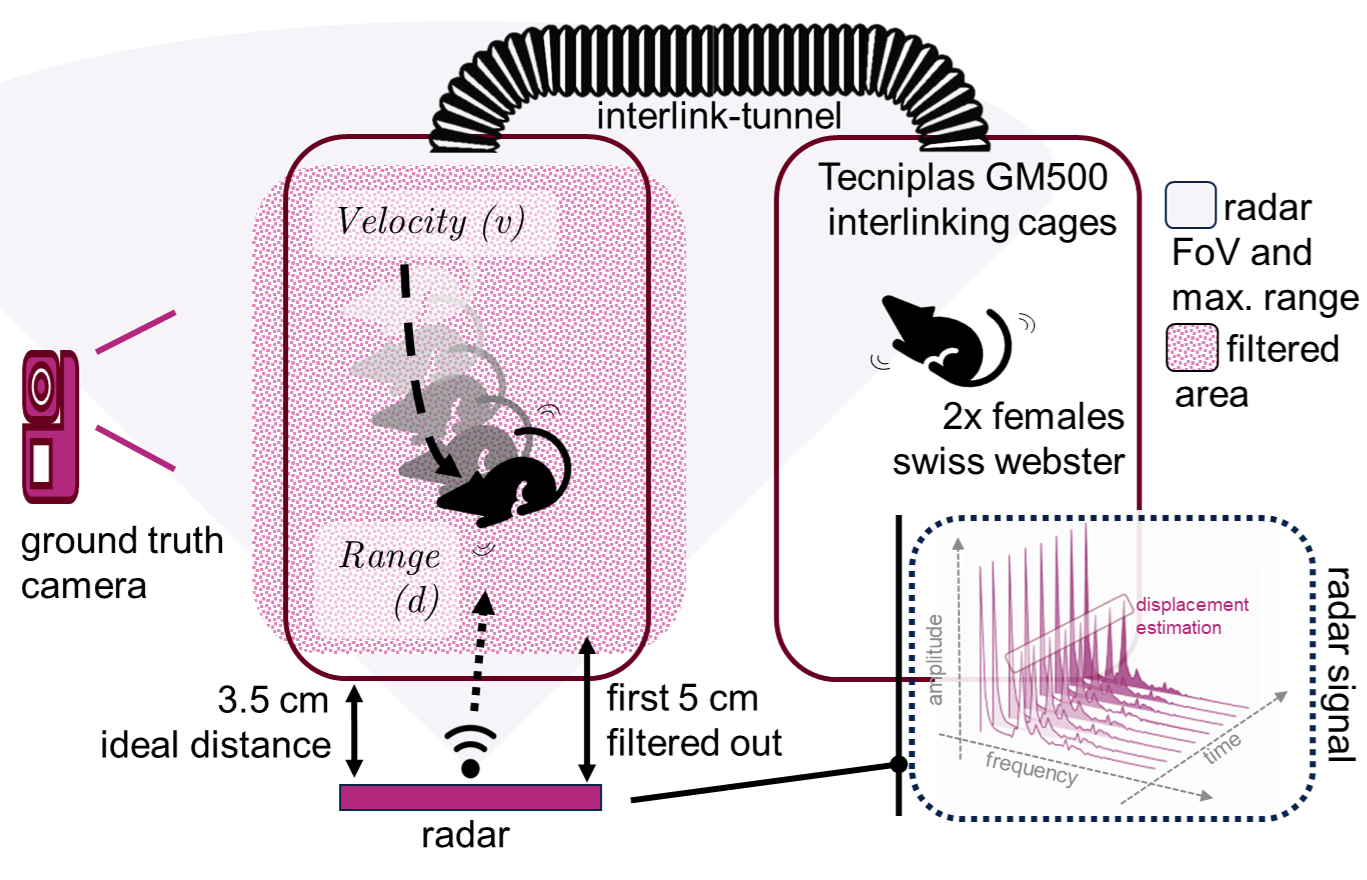}
    \caption{System setup overview for validation and in-vivo tests. The ideal distance is disccused in \Cref{sec:idealdistance}, velocity is defined in \Cref{eq:omega}-\Cref{eq:vres} and range $d$ in \Cref{eq:submillimeterdisplacements}. The \texttt{filtered area} highlights the area in which vital sing monitoring and tracking are active, signals coming outside this region are not considered.}
    \label{fig:overview}
\end{figure}

An overview of the system hardware and configuration is depicted in \Cref{fig:overview}, showing the radar position, distance and velocity estimation, other than the cage setting.

\subsection{Radar Hardware}
Infineon's Radar Baseboard MCU7 Plus, combined with the BGT60UTR13DAiP radar frontend, forms the selected radar platform. Designed for short-range sensing applications, the BGT60UTR13DAiP is a low-power, high-resolution FMCW radar IC operating in the \qty{60}{\giga\hertz} domain with a maximum bandwidth of  \qty{5.5}{\giga\hertz}. One transmitting and three receiving antennas are directly integrated on the back of the IC, arranged in an L-shape. For efficient data transfer, the radar baseboard includes a high-speed USB 2.0 interface, allowing real-time streaming to a host PC. The radar platform is shown in \Cref{fig:Höngg_experiment_setup}. %To facilitate development, Infineon provides the Radar Development Kit, which includes the Fusion GUI for intuitive configuration and visualization, along with SDKs for programming in Python, C, C++, and MATLAB.
\subsection{Radar FMCW Configuration}
The BGT60UTR13D radar sensor offers a high degree of configurability. For rodent monitoring, we are interested in both the overall movement of the mice and the imperceptible skin vibration caused by their heartbeats and breathing. Therefore, a single radar configuration would result is a sub-optimal measurement accuracy, not perfectly addressing both aspects. A high frame rate is desirable for vital sign signal processing; however, it comes at the cost of reduced velocity resolution due to the shorter frame time (\Cref{eq:vres}). To resolve this trade-off, we present two distinct radar configurations: one optimized for detecting coarse movements, referred to as the "Movement Configuration," and another designed for capturing fine movements, called the "Vital Sign Configuration". The BGT60UTR13D can be programmed in runtime switching between these two settings with negligible latency, both configurations are summarized in \Cref{tab:radar_config}. 
\subsubsection{Movement Configuration} Considering the maximum velocity of a mouse inside a cage, as discussed in \Cref{sec:mousephysiology} a value of \qty{185}{\micro\second} is derived for \(T_{c,r}\) using \Cref{eq:omega} yielding $v_{max} =$ \qty{6.7}{\meter\per\second}. To ensure sufficient velocity resolution for movement monitoring, 128 chirps per frame are used, resulting in $v_{res} =$ \qty{0.105}{\meter\per\second}, as determined by \Cref{eq:vres}. For appropriate range resolution, a bandwidth of \qty{5}{\giga\hertz} is used, resulting in a range resolution of \qty{3}{\cm}, as determined by \Cref{eq:submillimeterdisplacements}. Finally, the total frame rate resulted in \qty{40}{\hertz} and a maximum ranging distance of $d_{max}=~$\qty{3.8}{\meter}.
\subsubsection{Vital Sign Configuration} 
The authors in \cite{marty_frequency_2024} successfully demonstrated vital sign monitoring in humans using a frame rate of \qty{100}{\hertz}. Since the HR and the RR of a mouse are significantly higher compared to those of a human (\Cref{sec:mousephysiology}) the total frame rate for the Vital Sign Configuration is set to \qty{400}{\hertz}. Additionally, \cite{marty_frequency_2024} reported that the concept of "chirp accumulation" can enhance the signal-to-noise ratio. Based on this, 16 chirps per frame are used in Vital Sign Configuration in \Cref{tab:radar_config}. The same \qty{5}{\giga\hertz} bandwidth as in the Movement Configuration is used. With these settings, the distance resolution reaches \qty{2}{\micro\meter}, measured from the phase difference $\Delta d$ and limited by the noise floor modeled in \cite{marty_frequency_2024}.
\\

\begin{table}[t]
\centering
\caption{BGT60UTR13D configurations used for the experimental setup and the dataset collection. The BGT60UTR13D is configured in runtime depending on the Movement or Vital Sign measurement.}
\label{tab:radar_config}
\begin{tabular}{lcc}
\hline\hline
\textbf{Parameter}       & \textbf{Movement} & \textbf{Vital Sign} \\
\hline
Start frequency (\( f_{\text{start}} \)) & \qty{58}{\giga\hertz} & \qty{58}{\giga\hertz} \\
End frequency (\( f_{\text{end}} \)) & \qty{63}{\giga\hertz} & \qty{63}{\giga\hertz} \\
Bandwidth (\( B \)) & \qty{5}{\giga\hertz}  & \qty{5}{\giga\hertz}  \\
Frame rate (\( f \)) & \qty{40}{\hertz}     & \qty{400}{\hertz}     \\
ADC sampling rate (\( F_s \)) & \qty{2}{\mega\hertz}  & \qty{2}{\mega\hertz}  \\
% Ramp-up slope (\( S \)) & \qty{39}{\mega\hertz\per\micro\second} & \qty{78}{\mega\hertz\per\micro\second} \\
Number of antennas (\( N_a \))           & 3                   & 3                   \\
Number of chirps (\( N_c \))             & 128                 & 16                  \\
Number of samples (\( N_s \))            & 256                 & 128                 \\
\hline
Maximum distance (\( d_{\text{max}} \))    & \qty{3.8}{\meter}    & \qty{1.9}{\meter}    \\
Maximum velocity (\( v_{\text{max}} \))    & \qty{6.7}{\meter\per\second} & - \\
Distance resolution \( d_{\text{res}} \)  & \qty{3}{\centi\meter} & \qty{2}{\micro\meter}$^\diamond$ \\
Velocity resolution (\( v_{\text{res}} \))  & \qty{0.105}{\meter\per\second} & - \\
\hline\hline
\multicolumn{3}{l}{
$^\diamond$ Noise floor of $\Delta d$ characterized in \cite{marty_frequency_2024}.
}
\end{tabular}
\end{table}

%To implement these configurations, the Fusion GUI interface provided by Infineon was used to set the appropriate register values within the radar system.

 % better to start with QVar because IMO the 
\subsection{Ideal Distance to Cage}\label{sec:idealdistance}
As previously mentioned, our monitoring system has the sensor placed outside the home-cage to minimize in-vivo test contamination and for facilitating installation. Therefore, between the BGT60UTR13DAiP radar and the rodent there is a plastic polycarbonate layer with variable thickness, depending on the specific home-cage model. In the field of radar sensing, this problem can be modeled as a radome, a dome-shaped structure protecting radar equipment. In our work, we considered a plastic thickness $\in [0.182, 0.365, 0.729, 1.459]$ \qty{ }{\milli\meter}, with a relative permittivity of $\epsilon_r = 2.6$. 

\begin{figure}[t]
    \centering
    \includegraphics[width=0.99\columnwidth]{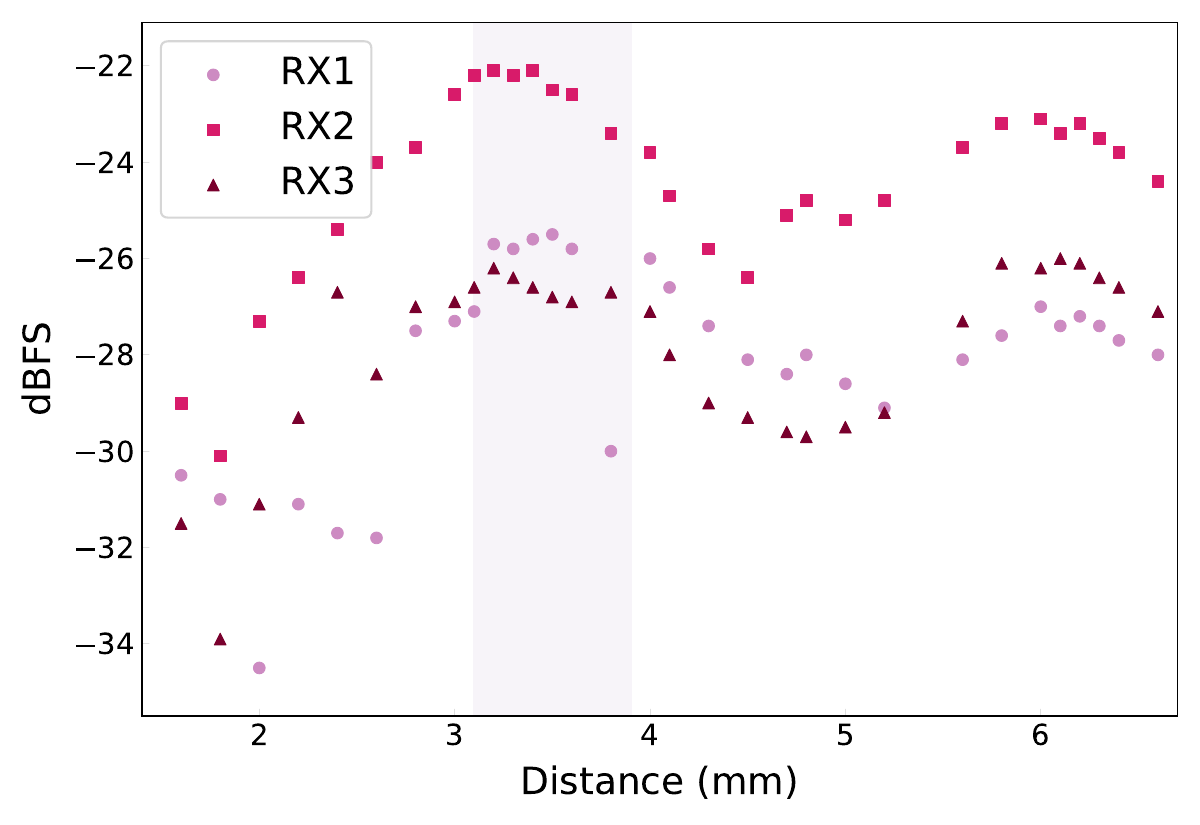}
    \caption{Filed analysis of signal reflection depending on the distance between the radar antennas and the cage plastic. Normalized signal strength is measured in dBFS (decibels relative to full scale). The optimal distance range for all the three antenna is highlighted between \qtyrange{3}{4}{\milli\meter}.}
    \label{fig:radome}
\end{figure}

The radome will impact on the radar performance reducing the signal strength of the detected radar targets; in particular, the change in the $\epsilon_r$ between air and radome causes the radar signal to be partially reflected when entering and also when exiting the radome. The magnitude of the reflection  rises with an increasing relative permittivity value of the radome material. While the literature suggests the distance between the BGT60LTR11AIP and the radome to be half the free-space wavelength ($\lambda/2 \simeq$ \qty{2.5}{\milli\meter} at \qty{60}{\giga\hertz}) or multiples of that, practical deployment adds more challenges. For example, the variable radome thickens and the frequency sweep between \qtyrange{58}{63}{\giga\hertz} play a role in generating non-ideal conditions. Therefore, we modeled our setup in analogy with the Fabry Perot approach\footnote{BGT60LTR11AIP radome design guide, Infineon AN091 611}, with a field carachterization. The optimal radar-radome distance resulted to be in the range \qtyrange{3}{4}{\milli\meter} (center at \qty{3.5}{\milli\meter}) and its multiple.  \Cref{fig:radome} reports the received power (normalized to \qty{0}{dBFS}) with variable radome-radar distance for the three RX antennas. 
Notably, for practical reasons a distance of \qty{3.5}{\milli\meter} is sub-optimal during the installation of the radar on the Tecniplast home-cage, therefore a $10\times$ multiple is selected, resulting in \qty{3.5}{\cm}.

\begin{figure}[t]
    \centering
    \includegraphics[width=0.99\columnwidth]{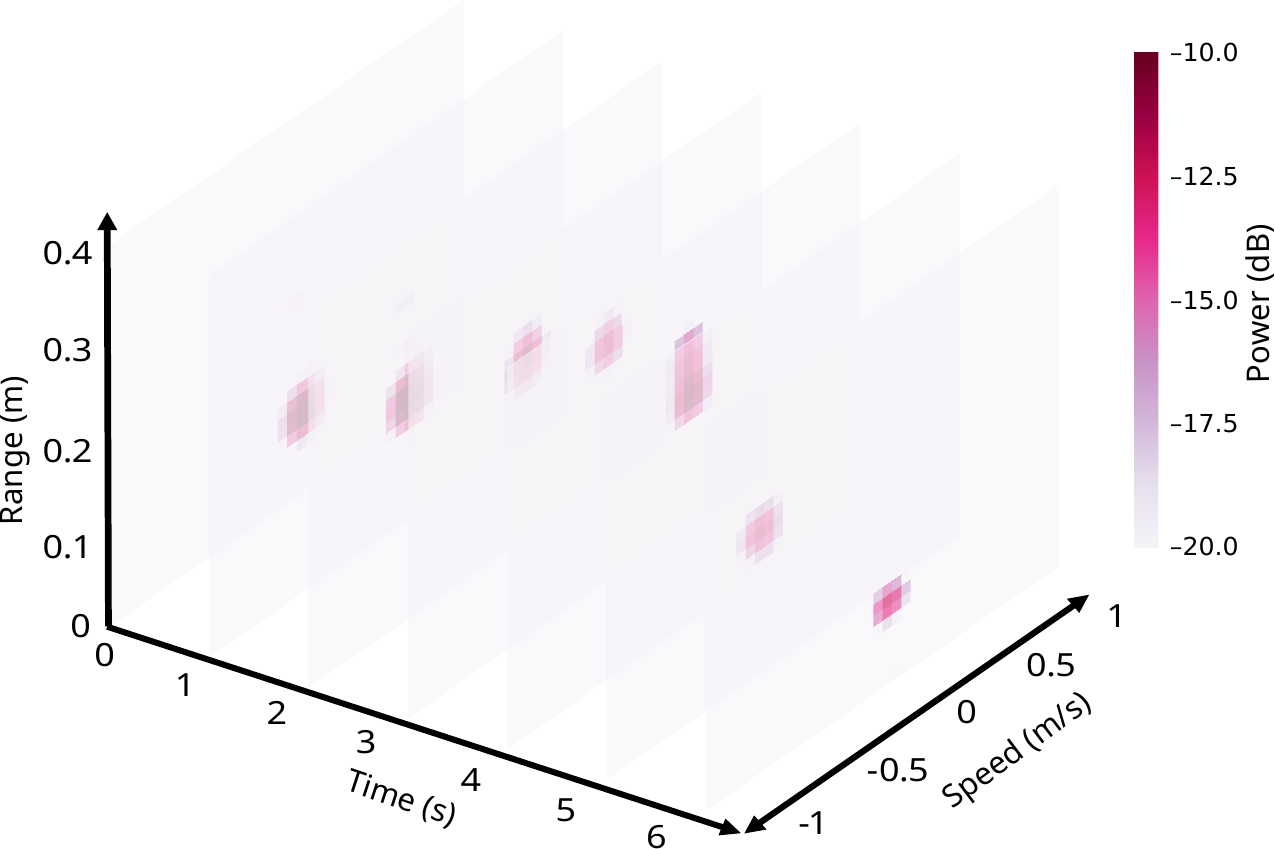}
    \caption{Ranging of a puppet target over time, represented as a series of one range-doppler maps per second. The sequence is recorded as part of the validation process in controlled conditions during a movement segment. The intense color represent the target.}
    \label{fig:range_doppler_time}
\end{figure}

\subsection{Movement Validation}\label{sec:calibrationmovement}
\begin{figure}[t]
    \centering
    \includegraphics[width=0.99\columnwidth]{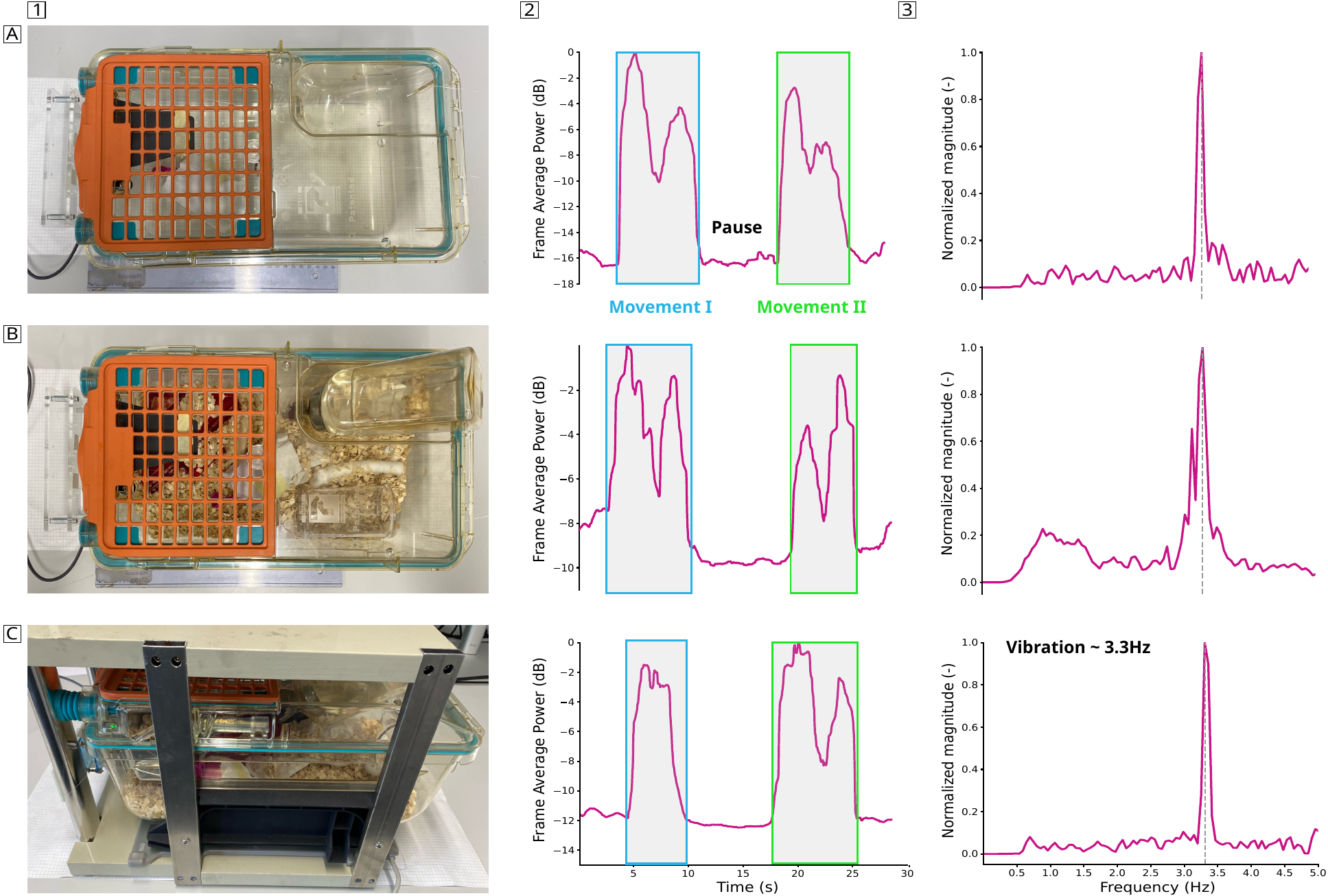}
    \caption{A. Empty cage. B. Cage filled with all internal elements. C. Filled cage complete of external metallic rack support. 1. Cage conditions. 2. Radar's average frame power during movement validation. In the blue rectangles the first movement segment and in the green rectangles the second movement segment. 3. FFT of the target vibration signal sensed with the radar. The gray dashed line indicates the frequency of the FFT's peak, corresponding to the measured target vibration frequency.} 
    \label{fig:movement_vibration}
\end{figure}
Before testing our system in the field, a comprehensive validation is conducted in controlled conditions. Throughout this process, we use a puppet to simulate the rodent's behavior and its body. The puppet is made out of Agar, a jelly-like substance consisting of polysaccharides obtained from the cell walls of red algae, primarily from "ogonori" and "tengusa", mimicking biological properties of animal tissue with the ability to simulate tissue conductivity and permittivity \cite{kruljac_challenges_2024,marty_investigation_2023}. The puppet produced has a quasi-cylindrical shape with a dimension of \qtyproduct{7 x 5 x 5}{\cm}.

As mentioned in \cref{sec:cage_env}, a rodent's cage is a challenging environment for radar sensing. Therefore, it is necessary to evaluate the effectiveness of our system in the presence of all objects and materials commonly present. These include cage bedding (burrowing substrate) - which is generally made out of wood shavings, or wheat/rice straw - wooden and plastic objects, food, water and their respective containers. Notably, the metallic structure surrounding the cage could also be a source of noise. Therefore, the validation is repeated in three subsequent levels of complexity, which can be observed in \Cref{fig:movement_vibration}.1: an empty cage, a cage with all internal elements, and a full-setup cage complete of external metallic rack support. The movement detection and ranging capabilities of our system are validated through \qty{30}{\second} recordings alternating movement and stationary segments. The puppet is moved with nylon strings along the full length of the cage and back. During each recording, the movement segments is performed roughly between \qtyrange{5}{10}{\second} and \qtyrange{20}{25}{\second}. The puppet is stationary for the remaining time. As described in \Cref{sec:dsp}, we use the L2 norm (frame average power) of the radar's range-doppler map to calculate the target's movement level. Furthermore, we can track the target's distance over time via subsequent range-doppler frames during one of the movement segments, as exemplified in \Cref{fig:range_doppler_time}.

Results of the movement detection validation are shown in \Cref{fig:movement_vibration}.2, in which the movement segments are highlighted in blue and green. These plots clearly confirm that the system is consistently capable of detecting the puppet movement at any level of cage complexity. As soon as the puppet starts to move, the frame power increases and it reaches its peak when the puppet is moving the most - at its maximum velocity. Upon reaching the back of the cage, the puppet slows down and the frame power starts decreasing. Then, the puppet is pulled back to the near side of the cage and consequently the frame power rises again. Note how the frame power arrives to a second peak before lowering when the puppet is finally stationary. This initial validation also demonstrate the adequacy of the frame power as a simple proxy for estimating rodents movement levels in the cage environment.
\Cref{fig:range_doppler_time} displays the ranging and localization ability of the system during one of the movement segments. The puppet is initially near the radar at around \qtyrange{5}{7}{\cm}, then it is dragged to the back end of the cage, \qtyrange{35}{37}{\cm} far from the radar. Finally, the puppet is pulled back towards the radar again at \qtyrange{5}{7}{\cm}. Therefore, \Cref{fig:range_doppler_time} demonstrates how we  can measure the target location throughout its movement along the cage and track it over time.

\subsection{Vital Sign Validation }
\label{sec:calibrationrespiration}
Following the rationale introduced in the previous section, an evaluation of our system's ability to measure rodent's vital signals is also performed. 
The system records \qty{20}{\second} of the puppet vibrating at a known frequency; it is gently poked at a rate of \qty{200}{bpm} with a diapason, generating sub-millimeter oscillations. This frequency is chosen to simulate the RR, as it falls within the expected range indicated in \Cref{sec:mousephysiology}. The vibration mimics the compression and expansion of the rodent's chest due to pulmonary respiration cycle, while the puppet is placed at the center of the cage and remains stationary for the whole duration of the recording. The experiment is repeated in the three cage conditions presented earlier. For each recording, we follow the signal processing steps described in \Cref{sec:dsp} to measure the vibration frequency of the target. The results of the vital signal validation are reported in \Cref{fig:movement_vibration}.3. The FFT peak matches the actuator frequency of \qty{200}{bpm} ($\sim$\qty{3.3}{\hertz}), showcasing the system's ability to accurately measure the rodent vital sign  inside the cage. Moreover, this result is consistent across all cage conditions. In the case of an empty cage, the peak is very sharp and there are no fluctuations or secondary peaks at other frequencies. For the recording with all internal elements, a component of low-frequencies and a secondary peak can be observed in \Cref{fig:movement_vibration}.3B. %These unwanted effects could arise as a consequence of the increased clutter in the cage. However, the test in the most complete cage condition shows once again a single and sharp peak. Therefore, the former observations are likely a simple result of empirical variability or inaccuracies of the actuator, and not related to the cage conditions.

\section{Experimental Setup \& Dataset Collection }
\label{sec:setup}
To validate the proposed radar platform for rodent monitoring, in-vivo experiments are conducted in the ETH Phenomics Center (EPIC), which is a state-of-the-art research institute at ETH Zürich  specialized in experimental laboratory animal research with a focus on molecular health sciences. 

\begin{figure*}[t]
    \centering
    \includegraphics[width=1.99\columnwidth]{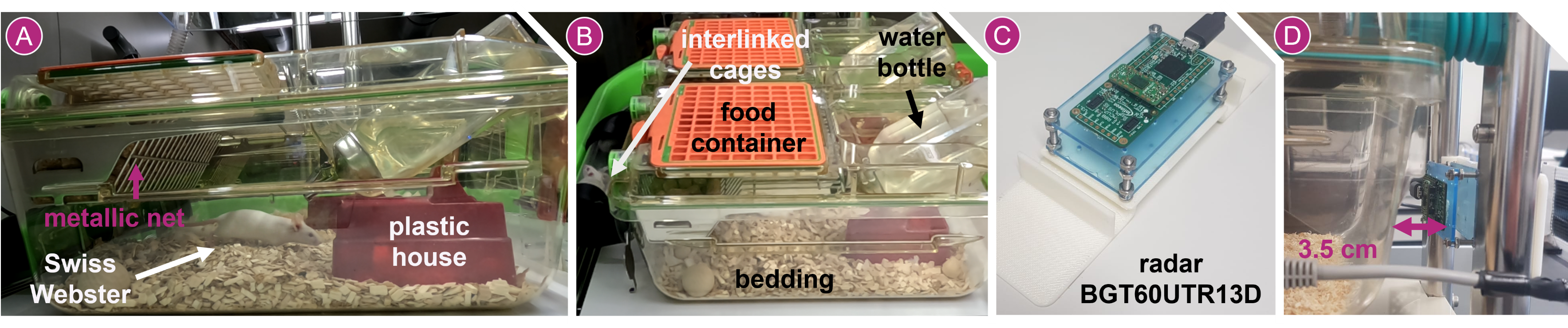}
    \caption{Setup for in-vivo experiments at ETH EPIC center. The two interlinked cages (model GM500 from Tecniplas) are mounted on a local changing station, allowing both mice to freely access either cage at any time during the experiment. Occasionally, the mice voluntarily separated themselves between the two cages, enabling "single-mouse" experiments to be conducted. The radar is placed at \qty{3.5}{\cm} to the cage, connected to a local workstation for data collection and synchronization with the image-based ground-truth. (A) Lateral view of the cage environment with metallic nets, internal plastic and the Swiss Webster; (B) Cages top view; (C) The BGT60UTR13DAiP radar electronic board used for the in-vivo experiments; (D) Ideal distance to cage technicality. } 
    \label{fig:Höngg_experiment_setup}
\end{figure*}

Mice are highly social animals, and housing them individually can negatively impact their welfare. In Switzerland, there are strict regulations that prohibit single housing for mice unless there are exceptional reasons\footnote{Swiss Animal Welfare Ordinance, Art. 119}. As described in \Cref{sec:dsp}, multi-mouse monitoring is expected to be significantly more complex than single-mouse monitoring. To enable initial experiments also with a single mouse, a workaround is developed. An innovative interlinking cage prototype is developed with the support of Tecniplast, which enables quasi-single mouse housing while complying with Swiss animal welfare regulations (see \Cref{sec:mousephysiology}). The cage setup is shown in \Cref{fig:Höngg_experiment_setup}, where two neighboring cages are connected by a tube, allowing the mice to arbitrarily access both cages at any time. The radar was placed as described in \Cref{sec:idealdistance} in front of one of the cages that we call the "monitoring cage". This approach allows for occasional single-mouse monitoring as well as multi-mice monitoring. A  camera is included into the experimental setup, serving as ground truth to derive the location and movement/state of the mice. Moreover, close-range camera recordings, combined with posterior video analysis, allowed for the derivation of ground truth for the mouse RR. However, due to the complete non-invasive approach of our tests, there is not a reference for the HR, which is only qualitatively evaluated against expected values.

The experimental protocol is structured as follows:
whenever the mice are in an active state (e.g., occasionally walking through the cage, moving in place, eating), a \qty{3}{\minute} radar measurement is initiated using the movement configuration in \Cref{tab:radar_config}. When the mice began to become sleepy, showing no to minimal stationary movement, \qty{3}{\minute} radar measurements are then started using the vital sign configuration. We repeat the procedure several times to capture  four  scenarios:
\begin{enumerate*}[label=(\roman*), font=\itshape]
    \item movement - single mouse,
    \item movement - two mice,
    \item vital sign - single mouse,
    \item vital sign - two mice.
\end{enumerate*}
"Single mouse" refers to time intervals during the experiment when the mice are distributed across two interlinked cages, with only one mouse present in monitoring cage. "Two mice," on the other hand, refers to intervals when both mice are present in the monitoring cage.

\section{In-Vivo Evaluation}
\label{sec:results}
The four in-vivo test scenarios are processed as described in \Cref{sec:dsp} to extract the RR, HR,  movement classification, and distance estimation. The chapters below cover the results of the four in-vivo experiments.

\subsection{Rodents for In-Vivo Evaluation}
\label{sec:mousephysiology}
Mice (females, Swiss Webster) are maintained in a DVC60 cage system (GM500 interlinking cages, Tecniplast), in a temperature- and humidity-controlled facility on a \qty{12}{\hour} light cycle with food and water ad libitum. The Tecniplast's home-cage is made out of transparent polycarbonate, where mice are housed in two per two interlinked cage. All procedures are carried out in accordance to Swiss cantonal regulations for animal experimentation\footnote{\url{www.blv.admin.ch/blv/en/home/tiere/tierversuche.html}}. Regarding the Swiss Webster characteristic, it is know that a laboratory mouse kept in a cage can reach a maximum velocity of approximately $\approx$\qty{3.5}{\meter\per\second} \cite{garland_maximal_1995}. Its RR typically ranges from \qtyrange{150}{300}{bpm} breaths per minute, while its HR falls within \qtyrange{500}{700}{bpm} \cite{mattson_comparison_2001}. Values that are used as reference to define the optimal radar settings, as these factors are typically influenced by the test environment, age, and the animal's physiological state.

The magnitude of cardiac-induced chest movement in Swiss Webster females is typically within tens of micrometers \cite{jones_vivo_2018}. The study in \cite{jones_vivo_2018} uses in-vivo imaging and piezo sensors to quantify in-plane chest wall displacement during the cardiac cycle.
Results define the in-plane chest displacement at \qty{32.4(31.6)}{\micro\meter}, and the out-of-plane (axial) displacement at \qty{3.9(4.7)}{\micro\meter}. These measurements reflect the mechanical movement at the chest surface and myocardium. Typical mouse chest displacement during respiration is in the range of \qtyrange{0.1}{0.5}{\milli\meter}, values related with the displacement between expiration and inspiration in non-anesthetized, healthy mice \cite{chang_synchrotron_2015}.

Regarding the environmental material, other than the cage polycarbonate, 
food and bedding must be considered. Mice are standardly housed on $\sim$\qty{140}{\gram} of SAFE ASPEN bedding material. Vitamin-fortified autoclaved food is from KLIBA NAFAG, similar to the autoclaved water provided in \qty{400}{\milli\liter} water bottles.

\subsection{Movement - Single Mouse}
\label{sec:mov_single}
\begin{figure}[t]
    \centering\includegraphics[width=0.99\columnwidth]{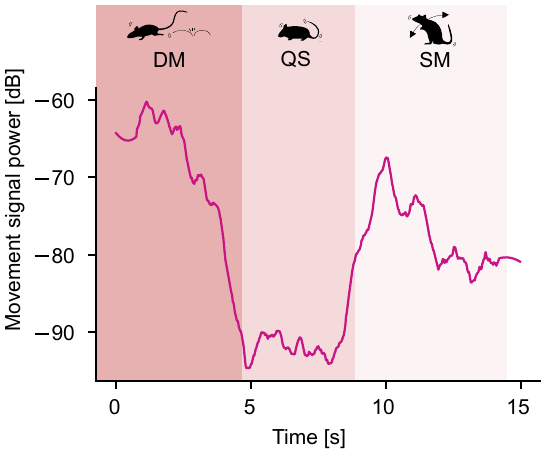}
    \caption{Single mouse inside the GM500 interlinking cage. This plot represents the movement level of the mouse, where the y-axis shows the signal power, and the x-axis represents time. The radar is configured in the Movement Detection mode. (DM = dynamic movement, QS = quasi-static, SM = stationary movement.)}
    \label{fig:movement_plot}
\end{figure}

In this subsection, the results of the movement level processing for the single mouse case are presented. The mice movement state can be categorized into three main groups: 
\begin{enumerate*}[label=(\roman*), font=\itshape]
    \item dynamic movement: mouse walking, running;
    \item static movement: mouse is stationary but has a ground level of activity e.g. mouse is cleaning itself, mouse is eating, mouse is sniffing around, etc.;
    \item quasi-static: lowest level of movement, baseline movement caused by breathing, during sleep mouse is in a quasi-static state.
\end{enumerate*} 
Results indicate that the proposed range-Doppler processing effectively detects mouse movements. The peak in the range-velocity map, corresponding to the moving mouse, is clearly emphasized relative to the surrounding background (see \Cref{fig:separate_mice_plot}). Moreover, the range-Doppler map for the two mice case looks comparable to the one for single mouse. The velocity of the mouse, as measured by the radar, remains predominantly within the range of $\pm$\qty{0.5}{\meter\per\second} which holds true with the estimated maximum velocity suggested in \Cref{sec:mousephysiology}. In \Cref{fig:movement_plot}, the movement and activity of the mouse (as described in \Cref{sec:dsp}) is derived from signal power over time. Between 0 and 5 seconds, the mouse moves dynamically inside the cage. From 5 to 8 seconds, the mouse remained in a quasi-static state. Finally, between 8 and 15 seconds, the mouse exhibited stationary movements. The movement signal power varies in accordance with the activity level of the mouse. Dynamic movement results in the highest power, followed by static movement and the quasi-static state. The power and spatial extent of the peak in the range-velocity plot depends also on the mouse’s proximity and posture relative to the radar, due to variations in the radar cross-section (RCS). This means that posture and proximity have a significant impact on the movement level estimation. This effect is particularly evident in the movement signal power plot between 2 seconds and 4 seconds, where the mouse moved away from the radar.

%From time to time multi path ghost objects appeared in the range-velocity plot that did not correspond to the actual range of the mouse when comparing to the camera ground-truth. 
    
%Occasionally, artifacts appeared in the radar signal due to the presence of a second mouse in the neighboring cage, which was also within the radar’s field of view.

% \begin{figure}[t]
%     \centering
%     \includegraphics[width=0.99\columnwidth]{fig/mouse_ranging.pdf}
%     \caption{\textcolor{red}{Alternative plot for the ranging}}
%     \label{fig:ranging_plot}
% \end{figure}

\begin{figure*}[t]
    \centering
    \includegraphics[width=1.99\columnwidth]{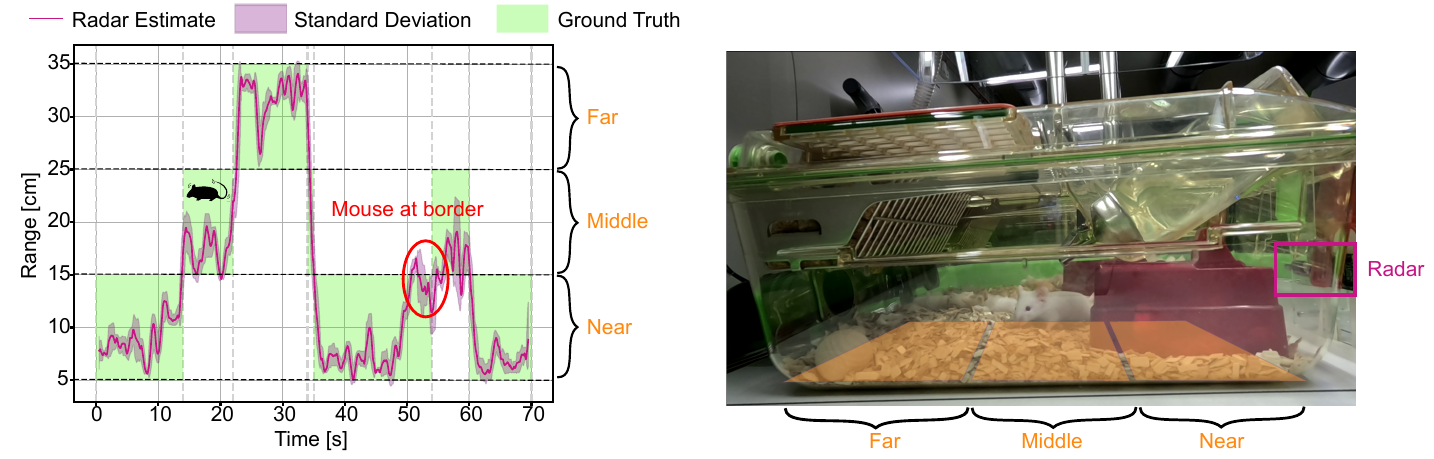}
    \caption{The left plot depicts the radar range estimation - nominal value and standard deviation. The light-green color indicates the ground truth zone where the mouse is actually present. On the right picture, a frame of the  ground truth video shows the division of the cage into three zones.} 
    \label{fig:mouse_ranging_and_cage}
\end{figure*}

\subsection{Movement - Two Mice}
\begin{figure}[t]
    \centering
    \includegraphics[width=0.9\columnwidth]{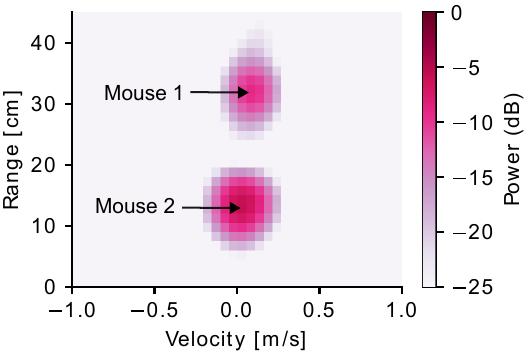}
    \caption{Range-velocity plot of two mice moving in the same GM500 cage. We clearly distinguish between two mice, allowing their movements to be individually identified and tracked.}
    \label{fig:separate_mice_plot}
\end{figure}

We repeated the analysis in \Cref{sec:mov_single} but only including recordings with two mice in the same cage. \Cref{fig:separate_mice_plot} shows the range-velocity plot of a specific frame, where both mice are visible as separate peaks, indicating their respective ranges and momentary velocities. \Cref{fig:separate_mice_plot} demonstrates the possibility of tracking the movement of several mice simultaneously.

It was assumed that movement level processing would allow, to some extent, an estimation of the number of mice moving inside the cage. For example, in the case of two mice, the general movement level baseline when both are moving could be expected to be significantly higher compared to when only one mouse is moving while the other remains stationary. However, analysis of the movement level plot shows that distinguishing between these scenarios is not reliable. As observed in the single-mouse experiments, the radar's movement level estimation is highly dependent on the posture and proximity of the mouse, further complicating assumptions when multiple mice are present.

\subsection{Ranging - Single Mouse}
\label{sec:ranging_single}
To assess the ranging capabilities of the radar a ground truth is derived from the video. For this scope, the cage is divided into three zones: "Near", "Middle" and "Far", as illustrated in \Cref{fig:mouse_ranging_and_cage}. Furthermore, only the data for single mouse experiments are considered. The radar estimate for the range is plotted over the time in \Cref{fig:mouse_ranging_and_cage}. It can be observed how the estimated range stays most of the time inside the ground truth zone colored in light-green. Around the time at \qty{54}{\second}, the mouse is exactly at the border of two zones with the head in the Middle zone and the tail in the Near zone leading to a range estimation ambiguity. Nevertheless, over the whole \qty{70}{\second} experiment in \Cref{fig:mouse_ranging_and_cage}, the correct zone got correctly identified with an accuracy of 93\%.

\subsection{Respiration - Single Mouse}
\label{sec:respiration_single}
\begin{figure}[t]
    \centering
    \includegraphics[width=0.99\columnwidth]{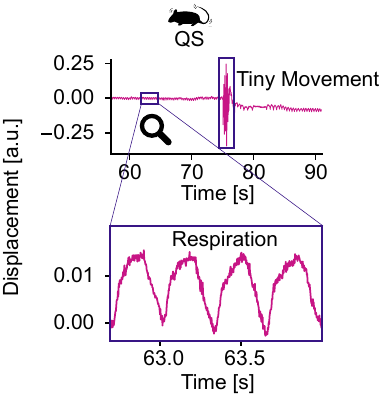}
    \caption{Analysis of the single-mouse respiration monitoring. The plot zoom-in a period of \qty{500}{\milli\second} to visualize the radar displacement of the chest pulmonary activity over time.}
    \label{fig:tiny_movement_and_respiration_plot}
\end{figure}

Analysis of the single-mouse respiration monitoring results demonstrate that the high sensitivity of the radar to fine movements enables precise tracking of both minute movements and respiration. \Cref{fig:tiny_movement_and_respiration_plot} shows the tiny change in displacement over time of a dozing mouse. The prominent signal fluctuation at around \qty{75}{\second} corresponds to a very small movement of the mous during the sleep phase. Zooming in on the signal at around \qty{63}{\second}, we can clearly observe the tiny oscillations that correspond to the respiration of the mouse by pulmonary activity. The estimated RR of the radar achieved an accuracy of 99.0\% over a time window of \qty{70}{\second}.

\subsection{Respiration - Two Mice}
The respiration estimation in the presence of multiple mice is aslo demonstrate with an identical setup of \Cref{sec:respiration_single}, but in this condition some exceptions and limitations exist.
During the experiments in which two mice are present, we notice that mice often sleep snuggled together. Depending on the posture, it can happen that one mouse is partly or completely in the radar shadow of the other. This makes it hard or impossible to measure the breathing signal in the radar shadow with the setup used for the scope of this work.
Another challenge arises when multiple mice share the same range bin, in this case the breathing information of multiple mice is reflected in one shared range bin and not as ideally spread over different range bins. Therefore, to distinguish between the individual RRs of multiple mice sharing the same range bin, the extracted phase signal containing the breathing signals is processed using FFT, with the goal to identify different peaks in the FFT corresponding to the individual RRs of the mice. However, in our experiments, the computed FFT only results in the peak that corresponded to the mouse closer to the radar, which is more exposed to the radar radiation. Several peaks in the FFT corresponding to different mice are not observed.

\subsection{Heart Beat - Single Mouse} 
Unfortunately, the heartbeat processing for the single mouse case have not resulted in any prominent periodic patterns indicative of a heartbeat. %Furthermore, since no ground truth was available, the heartbeat estimation could not be empirically evaluated.
Extensive signal processing is applied to the displacement signal to capture the heartbeat oscillations of the mouse. Despite employing several signal processing steps, such as DC offset removal and high-frequency processing \cite{marty_frequency_2024}, and ensuring optimal experimental conditions (e.g., positioning the radar module just below the cage, very close and facing the mouse's chest), no clear heartbeat oscillations are observed. %The SNR is probably not  enough for the BGT60UTR13DAiP to capture the tiny micro-meter movements of a mouse's heart beat.

\subsection{Limitations}
The proposed signal processing approach for ranging (\Cref{sec:ranging_single}) is expected to also work for multi-mouse monitoring. However, several additional steps are required, which are outside this paper scope. Currently, after deriving the range-power array, the processing selects the maximum range-power bin and identifies it as the range to the mouse. This approach becomes insufficient when multiple mice are present. Instead, a peak detection algorithm with distance-power compensation should be employed to identify the range to multiple mice. Furthermore, tracking the range over time using concepts such as a Kalman filter could enhance the robustness of the range estimates. 
%A limiting factor of the in vivo experiments with mice in Switzerland is the strict regulations, as mentioned in \Cref{sec:setup}. Despite Tecniplast’s efforts in developing a prototype for combined housing to enable quasi-single housing, the mice often stayed close together and slept cuddled up. This makes it challenging to acquire data from an individual sleeping mouse.

A significant challenge in multi-mice experiments is the increased complexity of the radar data processing, as unwanted effects such as multipaths become more pronounced, and additional phenomena, like shadowing, are introduced. Shadowing occurs when multiple mice align with the radar, and the mouse closer to the radar reflects all the signal, leaving no signal to be returned from the mice located in the shadow of the closer one. Consequently, monitoring the movement of multiple mice becomes challenging when using the range-velocity plot.

\section{Conclusion}
\label{sec:conclusion}

This paper presents an accurate contactless and non-invasive monitoring system for in-vivo rodent studies based on a low-power \qty{60}{\giga\hertz} FMCW radar. The system is designed to track animal position and classify activities while simultaneously extracting vital signs such as respiration rate and heart rate in freely moving rodents. Unlike conventional approaches that rely on invasive sensors, restraint, or frequent human handling, our radar-based method enables continuous monitoring directly in the home cage environment. 

Experimental validation demonstrated that the system can detect rodent activity and position with high reliability and estimate respiration rates within an error of less than $\pm$\qty{2}{bpm} compared to reference measurements. Heart rate monitoring is achieved only in controlled conditions, with a resolution sufficient to capture fluctuations in the \qtyrange{500}{600}{bpm} range typical of mice, although sensitivity decreases in highly active periods. The system maintained consistent respiration tracking in a standard cage for periods exceeding 2 hours. The radar operated with power consumption below \qty{5}{\milli\watt} in duty-cycled mode, confirming its suitability for long-term, battery-powered operation.  

Beyond the pioneering scientific contributions in electronics and information technology, this work has direct implications for the 3R principles. By eliminating the need for implants or handling, it significantly refines animal experimentation, reducing stress responses that can alter physiology and confound results. For example, handling alone has been shown to increase corticosterone levels by up to 30\% in laboratory mice~\cite{hurst_taming_2010}, while non-invasive home-cage monitoring reduces stress-induced variability and improves reproducibility~\cite{frohlich_editorial_2023}. The continuous and high-fidelity data acquisition also supports reduction, as fewer animals are required to obtain statistically valid datasets. Ultimately, this paper has the potential to push for a new standard in preclinical monitoring, improving both the quality of experimental data and the welfare of laboratory animals. Future work will focus on scaling the system for larger cohorts, integrating advanced AI models for stress and health assessment, and extending the framework to other species.

\bibliography{references}
\bibliographystyle{IEEEtran}

\end{document}